\documentclass[english,10pt,twocolumn,onespace]{IEEEtran}
\usepackage[T1]{fontenc}
\usepackage[latin9]{inputenc}
\usepackage{bm}
\usepackage{amsthm}
\usepackage{amsmath}
\usepackage{amssymb}
\usepackage{graphicx}

\makeatletter
 \let\oldforeign@language\foreign@language
 \DeclareRobustCommand{\foreign@language}[1]{%
   \lowercase{\oldforeign@language{#1}}}
\theoremstyle{plain}
\newtheorem{thm}{\protect\theoremname}
\theoremstyle{plain}
\newtheorem{lem}[thm]{\protect\lemmaname}
\theoremstyle{plain}
\newtheorem{cor}[thm]{\protect\corollaryname}

\usepackage{bm}
\usepackage{subfigure}
\usepackage{cite}
\usepackage[english]{babel}
\usepackage[T1]{fontenc}
\usepackage{algorithm}
\usepackage{algorithmic}
\vspace{-1in}

\makeatother

\usepackage{babel}
\providecommand{\corollaryname}{Corollary}
\providecommand{\lemmaname}{Lemma}
\providecommand{\theoremname}{Theorem}

\begin{document}

\title{Joint Channel and Queue Aware Scheduling for Wireless Links with
Multiple Fading States}

\author{Juan Liu$^{\dagger}$, Wei Chen$^{*}$, \IEEEmembership{Senior Member, IEEE},
and Khaled B. Letaief$^{\dagger}$, \IEEEmembership{Fellow, IEEE}
\\$^{\dagger}$Department of Electrical and Computer Engineering\\The
Hong Kong University of Science and Technology, Clear Water Bay, Hong
Kong\\$^{*}$Department of Electronic Engineering, Tsinghua University,
Beijing 100084, China\\ Email: eejliu@ust.hk,$\,$wchen@tsinghua.edu.cn,$\,$eekhaled@ust.hk}

\markboth{}{{\normalsize{}}}
\maketitle
\begin{abstract}
 In this work, we address the delay optimal scheduling problem for
wireless transmission with fixed modulation over multi-state fading
channels. We propose a stochastic scheduling policy which schedules
the source to transmit with probability jointly based on the buffer
and channel states, with an average power constraint at the transmitter.
Our objective is to minimize the average queueing delay by choosing
the optimal transmission probabilities. Using Markov chain modeling,
we formulate a power-constrained delay minimization problem, and then
transform it into a Linear Programming (LP) one. By analyzing its
property, we can derive the optimal threshold-based scheduling policy
together with the corresponding transmission probabilities. Our theoretical
analysis is corroborated by simulation results. 
\end{abstract}

\section{{\normalsize{}Introduction}}

Wireless networks accommodate various multimedia traffics with different
Qualities of Service (QoS) for mobile users. For high-speed real-time
applications, the average delay packets experience and energy efficiency
become more critical considerations \cite{2014_JGAndrews_5G}, when
data packets are delivered over time-varying wireless fading channels.
Therefore, it is important to schedule data transmissions to minimize
the average delay given precious system resources (e.g., average power
and energy). This naturally leads to a cross-layer design issue, since
the average power and delay are metrics of physical and Medium Access
Control (MAC) layers, respectively. 

In wireless networks, cross-layer design on power-efficient and delay-minimal
transmission has been an ever-present important issue \cite{1999_Collins_Cruz_Allerton,2002_Berry_Gallager_TIT,2013_RABerry_TIT}.
A cross-layer scheduling policy was firstly proposed in \cite{1999_Collins_Cruz_Allerton}
to transmit data over a discrete-time two-state channel under the
constraints of the average delay and peak transmitter power.  In
\cite{2002_Berry_Gallager_TIT}, Berry and Gallager considered a general
cross-layer model where the user's transmission power and data rate
were allocated according to the current buffer state and the channel
state in each slot. The asymptotic optimal power-delay tradeoff was
derived for the large-delay and small-delay regimes in \cite{2002_Berry_Gallager_TIT}
and \cite{2013_RABerry_TIT}, respectively. The existence of stationary
average delay optimal policy was shown and some structural results
were obtained in \cite{2003_AKumar_infocom}. In these works, the
cross-layer scheduling problems were formulated using the theory of
Markov decision processes and the optimal solutions were numerically
computed using the dynamic programming technique. 

Recently, different optimization techniques were applied to find the
optimal power/energy-efficient scheduling policies under various constraints
on rate, delay and maximum transmission power \cite{2013_ThoLeNgoc_TWC,2014_AShami_QoS,2002_AElGamal_TN,2007_WChen_IWCLD}.
Meanwhile, the attempts to find analytical solutions have been made
for pursuit of deep insights in protocol designs. In \cite{2002_AElGamal_TN},
an energy efficient scheduling problem was studied for transmitting
the packets within a certain amount of time. Through probabilistic
analysis, the authors derived the optimal offline scheduling algorithm
with an infinite delay constraint and Poisson data arrival. In \cite{2007_WChen_IWCLD},
we considered a cross-layer scheduling problem for a two-state wireless
channel. The transmission power is adjusted according to the channel
quality to achieve the target Bit Error Rate (BER). To exploit the
power efficiently, the source transmits one packet in each slot if
the channel state is ``good'', and otherwise transmits with probability
based on the buffer length. The optimal policy was to transmit based
on a critical threshold on the queue length.  

In this paper, we generalize the work in \cite{2007_WChen_IWCLD}
to the scenario where data packets of real-time traffics are delivered
over multi-state wireless fading channels. The transmission power
is adapted in response to the channel state and fixed modulation is
applied to reduce the complexity of the transceivers and delay jitter.
We propose a stochastic scheduling policy where the source transmits
with probability based on both the channel and buffer states. Using
Markov chain modeling, we formulate a power-constrained delay minimization
problem, and then transform it into a Linear Programming (LP) one.
By exploiting its properties, we are able to obtain an elegant closed-form
expression for the optimal solution, from which we can finally determine
the optimal probabilistic transmission parameters. We show that there
is a critical threshold imposed on the queue length associated with
each channel state and vice versa. For example, the transmitter of
the source is inactive if the data queue length is below the critical
threshold, and active otherwise. We also validate the optimality of
the proposed policy in \cite{2007_WChen_IWCLD} for a two-state wireless
channel. 

The rest of this paper is organized as follows. Section \ref{sec:System-Model}
introduces the system model and the stochastic scheduling scheme.
In Section \ref{sec:Problem_Formulation}, a discrete-time Markov
chain model is constructed and an LP problem is formulated. The optimal
scheduling policy is analyzed in Section \ref{sec:LP_problem}. Section
\ref{sec:Simulation_Results} demonstrates the simulation results
and Section \ref{sec:Conclusions} concludes this paper. Throughout
this paper, the proofs are omitted due to limited space.

\section{{\normalsize{}System Model\label{sec:System-Model}}}

\subsection{{\normalsize{}System Description}}

\begin{figure}[t]
\centering
\renewcommand{\figurename}{Fig.}

\includegraphics[width=0.45\textwidth]{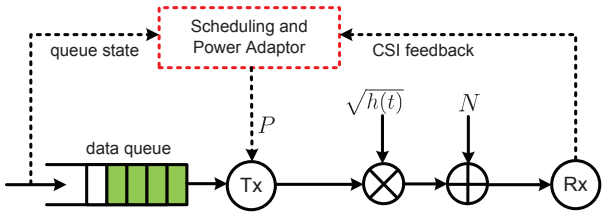}\protect\caption{System model. }
\label{fig:system_model}
\end{figure}

We consider a wireless link, where a source node transmits to its
destination over a time-varying channel. The source node employs a
buffer to store the data packets randomly generated from higher-layer
applications, as shown in Fig.$\,$\ref{fig:system_model}. The system
is assumed to be time-slotted.

Suppose that data packets arrive at the source buffer according to
a Bernoulli process \cite{2000_TGRobertazzi_book} with probability
$\alpha$. This simple yet widely adopted traffic model allows tractable
analysis \cite{2010_JY_Ulukus_tcom,2012_WChen_TNET}, and provides
insights for further study. The data buffer has a capacity of $Q$
($Q\in\mathbb{Z}^{+}$). $Q=\infty$ and $Q<\infty$ mean that the
buffer capacity is infinite and finite, respectively. Let $q[t]\in\mathcal{Q}=\{0,1,2,\cdots,Q\}$
be the number of backlogged data packets in the buffer at the end
of slot $t$, updated as
\begin{equation}
q[t]=\min\{q[t-1]+a[t],Q\}-v[t],
\end{equation}
where $a[t]$ and $v[t]$ denote the number of data packets arriving
and delivered in each time slot $t$, respectively. The probability
mass function of $a[t]$ can be expressed as $\Pr\{a[t]=1\}=\alpha,\,\Pr\{a[t]=0\}=1-\alpha$. 

\begin{figure}[t]
\centering
\renewcommand{\figurename}{Fig.}

\includegraphics[width=0.4\textwidth]{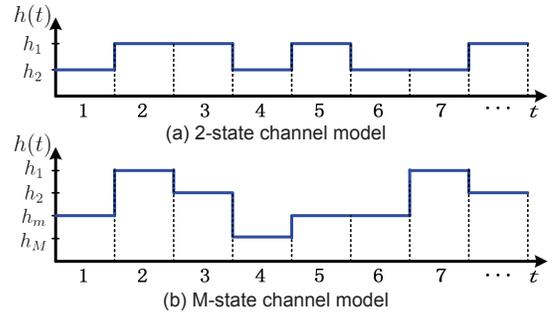}\protect\caption{Channel model. }
\label{fig:channel_model}
\end{figure}

The channel is assumed to experience $M$-state block fading, as
shown in Fig.$\,$\ref{fig:channel_model}. The channel state in slot
$t$ is denoted by $h(t)$ $(h(t)\in\mathcal{M}=\{1,\cdots,M\})$.
Assuming independent and identically distributed ($i.i.d.$) channel
fading model, we denote by $\eta_{m}$ the probability that the channel
stays at state $m$. The probability mass function of $h[t]$ is given
by $\Pr\{h[t]=m\}=\eta_{m}\,(m\in\mathcal{M})$, which satisfies $\sum_{m=1}^{M}\eta_{m}=1$.
 If the channel state is $m$, the source will transmit with power
$P_{m}$ to achieve a target BER. In practice, the transmission power
can be adjusted based on the received signal-to-noise ratio (SNR)
at the destination. Such Channel State Information (CSI) can be fed
back to the source over control channel. Intuitively, more power is
required to transmit one packet per slot when the received SNR is
smaller. Therefore, we naturally assume $P_{1}\leq P_{2}\leq\cdots\leq P_{M}$
in accordance with deteriorating channel conditions. In this sense,
$M$-state channel model is reasonable. Transmission schemes with
adaptive modulation and coding will be considered in our future work.
 

\subsection{{\normalsize{}Stochastic Scheduling}}

To improve the energy efficiency, the source is willing to wait for
better channel conditions, since it can thus spend less power on each
data transmission. However, the waiting time could be undesirably
long if the channel stays at very poor states for a long time. To
reduce the latency, the source may have to transmit its backlogged
data packets when the current channel state is not so good. This will
certainly cost more valuable power resource. Hence, there exists a
delay-power tradeoff in the wireless transmissions. 

Our objective is to find a scheduling policy that minimizes the average
queueing delay under the constraint of a maximum average transmission
power. To this end, we propose a stochastic scheduling scheme which
decides whether to transmit in slot $t$ according to the current
channel state $h(t)$ and data queue state $q[t-1]$. We define two
sets of probabilistic parameters: $\{g_{i,m}\}$ and $\{f_{i,m}\}$
($g_{i,m},f_{i,m}\in[0,1]$). Specifically, with $q[t-1]=i$ and $h(t)=m$,
if there is new data arrival in this slot, $i.e.$, $a[t]>0$, the
source node transmits one data packet with probability $g_{i,m}$
and holds from transmission with probability $1-g_{i,m}$, respectively;
If no new data packet arrives, $i.e.$, $a[t]=0$, it transmits with
probability $f_{i,m}$ and holds with probability $1-f_{i,m}$, respectively.

According to the proposed scheduling policy, the service process $v[t]$
depends on the queue status $q[t-1]$ and the arrival process $a[t]$,
as described below.
\begin{itemize}
\item Case 1: $q[t-1]=0$ and $h(t)=m$ 
\end{itemize}
In this case, the source transmits a newly arriving data packet with
probability $g_{0,m}$ in the current time slot $t$, and the service
process can be expressed as
\begin{equation}
v[t]=\begin{cases}
1\quad w.p.g_{0,m}, & a[t]=1,\\
0\quad w.p.(1-g_{0,m}), & a[t]=1,\\
0\quad w.p.1, & a[t]=0,
\end{cases}
\end{equation}
where $w.p.$ means 'with the probability'. 
\begin{itemize}
\item Case 2: $q[t-1]=i$ ($i>0$) and $h(t)=m$ 
\end{itemize}
In this case, the source transmits a packet with probability $g_{i,m}$
or $f_{i,m}$ depending on whether there is a new data arrival or
not. The service process can be expressed as
\begin{equation}
v[t]=\begin{cases}
1\quad w.p.g_{i,m}, & a[t]=1,\\
0\quad w.p.(1-g_{i,m}), & a[t]=1,\\
1\quad w.p.f_{i,m}, & a[t]=0,\\
0\quad w.p.(1-f_{i,m}), & a[t]=0.
\end{cases}
\end{equation}

\section{{\normalsize{}Problem Formulation\label{sec:Problem_Formulation}}}

\subsection{{\normalsize{}Markov Chain Model}}

\begin{figure}[t]
\centering
\renewcommand{\figurename}{Fig.}

\includegraphics[width=0.5\textwidth]{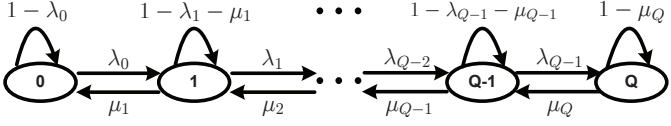}\protect\caption{Discrete-time Markov chain model of the buffer state. }
\label{fig:markov_chain_onefig}
\end{figure}

In our system, the queueing system can be modeled as a discrete-time
one-dimensional Markov chain, each state of which represents the buffer
status, as shown in Fig.$\,$\ref{fig:markov_chain_onefig}.

Let $\gamma_{i,j}=\Pr\{q[t+1]=j|q[t]=i\}$ denote the one-step transition
probability of the Markov chain, which is homogeneous by the scheme
description. The index $t$ can be omitted below if no confusion will
be caused. According to the case when one data packet newly arrives
while no transmission takes place, the transition probability that
the queue length is increased by one is obtained as
\begin{equation}
\lambda_{i}=\gamma_{i,i+1}=\alpha\sum\nolimits _{m=1}^{M}\eta_{m}(1-g_{i,m})\,(0\leq i\leq Q-1).\label{eq:lambda_i}
\end{equation}
When one data packet is transmitted with no new data arrival, the
transition probability that the queue length is decreased by one is
given by
\begin{equation}
\mu_{i}=\gamma_{i,i-1}=(1-\alpha)\sum\nolimits _{m=1}^{M}\eta_{m}f_{i,m}\,(1\leq i\leq Q).\label{eq:mu_i}
\end{equation}
Thus, the probability that the queue status remains the same can be
expressed as
\begin{equation}
\gamma_{i,i}=\begin{cases}
1-\lambda_{0}, & i=0,\\
1-\lambda_{i}-\mu_{i}, & 1\leq i\leq Q-1,\\
1-\mu_{Q}, & i=Q.
\end{cases}
\end{equation}
Let $\pi_{i}$ denote the steady-state probability that the data
queue length is equal to $i$. The steady-state probability vector
$\bm{\pi}=[\pi_{0}\,\pi_{1}\cdots\pi_{Q}]$ satisfies $\bm{\pi}\bm{P}=\bm{\pi}$
and $\bm{\pi}\bm{e}=1$. In particular, the local balance equation
at state $q[t]=i$ is given by
\begin{equation}
\pi_{i}\lambda_{i}=\pi_{i+1}\mu_{i+1},\,(0\leq i\leq Q-1).\label{eq:local_equation}
\end{equation}
Hence, the steady-state probability of the Markov chain can be computed
as
\begin{equation}
\pi_{0}=\left(1+\sum_{i=1}^{Q}\prod_{n=0}^{i-1}\frac{\lambda_{n}}{\mu_{n+1}}\right)^{-1},\,\pi_{i}=\pi_{0}\prod_{n=0}^{i-1}\frac{\lambda_{n}}{\mu_{n+1}}\,(i>0).\label{eq:prob_pi_i}
\end{equation}
Thus, given the transmission parameters $\{g_{i,m}\}$ and $\{f_{i,m}\}$,
we can compute the stationary distribution of the buffer state and
further analyze the system performance.

\subsection{{\normalsize{}Queueing Delay and Power Consumption }}

The system performance is measured in terms of the average queueing
delay and the average power consumption. By the Little\textquoteright s
law, the average queueing delay is related to the average buffer occupancy\cite{2002_Berry_Gallager_TIT},
and can be computed as 
\begin{equation}
\bar{D}=\frac{1}{\alpha}\sum\nolimits _{i=0}^{Q}i\pi_{i}.\label{eq:delay}
\end{equation}
Let $c[t]$ denote the transmission power in slot $t$. Let $\omega_{i,m}(x)=\Pr\{c[t]=x|q[t-1]=i,h(t)=m\}$
denote the conditional probability that the transmission power is
$c[t]=x$ $(x\in\{0,P_{m}\})$ given the data queue state $q[t-1]=i$
and the channel state $h(t)=m$. In this case, the source transmits
at a power $c[t]=P_{m}$ with probability $g_{i,m}$ if one data packet
newly arrives, and with probability $f_{i,m}$ if no data packet arrives,
respectively. Hence, conditioned on $q[t-1]=i$ and $h(t)=m$, the
probability that the transmission power is equal to $P_{m}$ can be
expressed as
\begin{equation}
\begin{split}\omega_{i,m}(P_{m})= & \begin{cases}
\alpha g_{0,m}, & i=0,\\
\alpha g_{i,m}+(1-\alpha)f_{i,m}, & 0<i\leq Q,
\end{cases}\end{split}
\end{equation}
and $\omega_{i,m}(0)=1-\omega_{i,m}(P_{m})$. Hence, the average power
is expressed as 
\begin{equation}
\begin{split}\bar{P} & =\sum_{i=0}^{Q}\pi_{i}\sum_{m=1}^{M}\eta_{m}\omega_{i,m}(P_{m})P_{m}\\
 & =\sum\limits _{m=1}^{M}(\alpha\sum\limits _{i=0}^{Q}\pi_{i}\eta_{m}g_{i,m}+(1-\alpha)\sum\limits _{i=1}^{Q}\pi_{i}\eta_{m}f_{i,m})P_{m}.
\end{split}
\label{eq:av_power}
\end{equation}

The event of packet loss occurs when there is one new data arrival
while the data buffer is full. Thus, the probability of buffer overflow
can be given by
\begin{equation}
p_{loss}=\Pr\{q[t]=Q\}\Pr\{a[t+1]=1|q[t]=Q\}=\alpha\pi_{Q}.\label{eq:packet_loss}
\end{equation}
Note that the average queueing delay can be appropriately defined
by (\ref{eq:delay}) only when the event of buffer overflow does not
take place. From (\ref{eq:packet_loss}), the packet loss probability
is zero if $\pi_{Q}$ is zero. This happens in two cases: 1) the buffer
capacity is infinite with $Q\rightarrow\infty$ and the queueing system
is stable; 2) the parameter $\lambda_{Q-1}$ is set to zero for a
finite buffer with $Q<\infty$. We will discuss the optimal scheduling
under the assumption that no buffer overflow occurs.

\subsection{{\normalsize{}Optimization Problem}}

 In this work, we aim to study the optimal scheduling policy which
minimizes the average delay $\bar{D}$ subject to the average power
constraint $\bar{P}\leq p_{max}$ by determining the optimal transmission
parameters $\{g_{i,m}^{*}\}$ and $\{f_{i,m}^{*}\}$. To this end,
we formulate an optimization problem as 
\begin{equation}
\begin{split}\min_{\{g_{i,m},f_{i,m}\}} & \quad\bar{D}=\frac{1}{\alpha}\sum\limits\nolimits _{i=0}^{Q}i\pi_{i}\\
s.t. & \begin{cases}
\bar{P}\leq p_{max}, & (a)\\
\pi_{i}\lambda_{i}=\pi_{i+1}\mu_{i+1},0\leq i<Q, & (b)\\
\sum\limits\nolimits _{i=0}^{Q}\pi_{i}=1, & (c)\\
g_{i,m},f_{i,m}\in[0,1],\:\forall i\in\mathcal{Q},m\in\mathcal{M}, & (d)
\end{cases}
\end{split}
\label{eq:non-linear-opt-problem}
\end{equation}
where the constraint (a) is the maximum average power constraint,
the constraints (b) and (c) stem directly from the property of the
Markov chain, and the constraint (d) points out the range of the probabilistic
parameters $\{g_{i,m}\}$ and $\{f_{i,m}\}$. Note that the average
power $\bar{P}$ (c.f. (\ref{eq:av_power})) and the steady-state
probabilities $\{\pi_{i}\}$ (c.f. (\ref{eq:lambda_i}), (\ref{eq:mu_i}),
(\ref{eq:prob_pi_i})) are non-linear functions of the parameters
$\{g_{i,m}\}$ and $\{f_{i,m}\}$. Therefore, it is rather difficult
to solve the above optimization problem (\ref{eq:non-linear-opt-problem}).
Motivated by the methods applied in \cite{2010_JY_Ulukus_tcom,2012_WChen_TNET},
we will transform the optimization problem (\ref{eq:non-linear-opt-problem})
into an LP problem, and exploit its special structure to analyze the
globally optimal solution in the next section.

\section{{\normalsize{}Analysis of Delay Optimal Scheduling\label{sec:LP_problem}}}

\subsection{{\normalsize{}LP Problem Formulation}}

To formulate an LP problem, we introduce a set of new variables
$\{y_{i,m}\}$ as: 
\begin{equation}
y_{i,m}=\pi_{i}g_{i,m}+\xi\pi_{i+1}f_{i+1,m},\label{eq:def_yim}
\end{equation}
where $\xi=\frac{1-\alpha}{\alpha}$. The variable $y_{i,m}$ can
be interpreted as the conditional probability that the queue state
is equal to $i$ after one data transmission over the wireless channel
with state $m$. By substituting (\ref{eq:lambda_i}) and (\ref{eq:mu_i})
into the local balance equation (\ref{eq:local_equation}), we have
$\pi_{i}\alpha\sum_{m=1}^{M}\eta_{m}(1-g_{i,m})=\pi_{i+1}(1-\alpha)\sum_{m=1}^{M}\eta_{m}f_{i+1,m}$,
from which we can further obtain
\begin{equation}
\pi_{i}=\sum\nolimits _{m=1}^{M}\eta_{m}y_{i,m}.\label{eq:prob_pi_yim}
\end{equation}
In Lemma \ref{lem:delay_power}, we show that the average delay $\bar{D}$
and power $\bar{P}$ are both linear functions of the variables $\{y_{i,m}\}$.
\begin{lem}
\label{lem:delay_power}The average delay $\bar{D}$ and the average
power consumption $\bar{P}$ can be expressed as 
\begin{equation}
\bar{D}=\frac{1}{\alpha}\sum_{m=1}^{M}\sum_{i=0}^{Q}i\eta_{m}y_{i,m},\,\bar{P}=\alpha\sum_{m=1}^{M}\sum_{i=0}^{Q}\eta_{m}P_{m}y_{i,m}.\label{eq:av_delay_v2}
\end{equation}

\end{lem}
As a result, we can transform the problem (\ref{eq:non-linear-opt-problem})
into an LP problem as follows:
\begin{equation}
\begin{split}\min_{y_{i,m}} & \quad\bar{D}=\frac{1}{\alpha}\sum_{i=0}^{Q}\sum_{m=1}^{M}i\eta_{m}y_{i,m}\\
s.t. & \begin{cases}
\bar{P}=\alpha\sum\limits _{i=0}^{Q}\sum\limits _{m=1}^{M}\eta_{m}P_{m}y_{i,m}\leq p_{max}, & (a)\\
\sum\limits _{i=0}^{Q}\sum\limits _{m=1}^{M}\eta_{m}y_{i,m}=1, & (b)\\
0\leq y_{i,m}\leq\sum\limits _{n=1}^{M}\eta_{n}y_{i,n}+\xi\sum\limits _{n=1}^{M}\eta_{n}y_{i+1,n}, & (c)\\
\forall i\in\mathcal{Q},m\in\mathcal{M}.
\end{cases}
\end{split}
\label{eq:LP_problem}
\end{equation}
In (\ref{eq:LP_problem}), the power constraint (a) and the normalization
constraint (b) are derived directly from the constraints (a) and (c)
of (\ref{eq:non-linear-opt-problem}). From (\ref{eq:def_yim}) and
(\ref{eq:prob_pi_yim}), we can derive the constraint (c) by varying
the probabilistic parameters $\{g_{i,m}\}$ and $\{f_{i,m}\}$ within
their range $[0,1]$. The optimal solution and the optimal value of
(\ref{eq:LP_problem}) are denoted by $\{y_{i,m}^{*}\}$ and $\bar{D}^{*}$,
respectively.

\subsection{{\normalsize{}Structure of the Optimal Solution}}

We first consider the case when the maximum average power $p_{max}$
is sufficiently large such that the source is able to transmit whenever
its queue is not empty. This means that the power constraint (\ref{eq:LP_problem}.a)
can be omitted. In this case, we can obtain the minimum average delay
$\bar{D}^{*}=\frac{1}{\alpha}\sum\limits\nolimits _{i=0}^{Q}i\sum_{m=1}^{M}\eta_{m}y_{i,m}^{*}=0$
by setting $y_{0,m}^{*}=1$ and $y_{i,m}^{*}=0\,(\forall i)$, which
satisfies the constraints (\ref{eq:LP_problem}.b) and (\ref{eq:LP_problem}.c).
Accordingly, we obtain the power threshold as $P_{th}=\alpha\sum_{m=1}^{M}\eta_{m}P_{m}$.

Then, we focus on the case when the power constraint (\ref{eq:LP_problem}.a)
becomes tight, $i.e.$, $\bar{P}=p_{max}<P_{th}$. By exploiting
the property of the LP problem (\ref{eq:LP_problem}), we can present
the structure of the optimal solution in the following theorem.
\begin{thm}
\label{thm:optimal_solution_structure}The optimal solution to (\ref{eq:LP_problem})
has a threshold structure. That is, there exists a threshold on the
data queue length $i_{m}^{*}$ associated with each channel state
$m$ such that \textup{\emph{$0=i_{1}^{*}\leq\cdots\leq i_{M}^{*}$
and the optimal solution takes the form as}}\textup{ 
\begin{equation}
y_{i,m}^{*}=\begin{cases}
0, & i<i_{m}^{*}-1,\\
\sum\limits _{n=1}^{M}\eta_{n}y_{i,n}^{*}+\xi\sum\limits _{n=1}^{M}\eta_{n}y_{i+1,n}^{*}, & i\geq i_{m}^{*},
\end{cases}\label{eq:optimal_yim_relationship}
\end{equation}
}And there is at most one $\tilde{m}\in\{2,\cdots,M\}$ such that
$y_{i_{m}^{*}-1,m}^{*}>0$ may hold, and $y_{i_{m}^{*}-1,m}^{*}=0$
for $m\in\{2,\cdots,M\}-\{\tilde{m}\}$.\end{thm}
\begin{cor}
\label{cor:larger_than_im}$y_{i,m}^{*}=0$ for all $i>i_{M}^{*}$
and $m\in\mathcal{M}$. 
\end{cor}
For all $i>i_{M}^{*}$ and $m$, we have $y_{i,m}^{*}=0$ and $\pi_{i}^{*}=\sum_{m=1}^{M}\eta_{m}y_{i,m}^{*}=0$.
This means that the length of the packet queue never exceeds the threshold
$i_{M}^{*}$. Therefore, no packet loss will be induced, $i.e.$,
$p_{loss}=0$, as long as the finite queue capacity $Q$ is larger
than $i_{M}^{*}$. 
\begin{figure}[t]
\centering
\renewcommand{\figurename}{Fig.}

\includegraphics[width=0.45\textwidth]{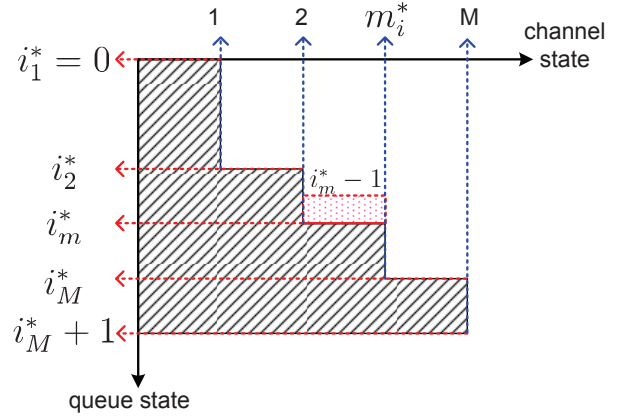}\protect\caption{The structure of the optimal solution: double threshold based policy. }
\label{fig:threshold_strucutre}
\end{figure}

As demonstrated by Fig.$\,$\ref{fig:threshold_strucutre}, the optimal
solution to problem (\ref{eq:LP_problem}) has a double threshold
structure:

1) There exists a threshold on the data queue length $i_{m}^{*}$
associated with each channel state $m$;

2) And there is a threshold imposed on the channel state $m_{i}^{*}$
given the queue state $i$. 

As shown by the shadow area in Fig.$\,$\ref{fig:threshold_strucutre},
we have $y_{i,m}^{*}=\pi_{i}^{*}+\xi\pi_{i+1}^{*}>0$ for $i_{m}^{*}\leq i\leq i_{M}^{*}$
or $m\leq m_{i}^{*}$, and otherwise $y_{i,m}^{*}=0$ for $i<i_{m}^{*}$
or $m>m_{i}^{*}$. The dotted area shown in Fig.$\,$\ref{fig:threshold_strucutre}
means that $0\leq y_{i,m}^{*}<\pi_{i}^{*}+\xi\pi_{i+1}^{*}$ for $i=i_{\tilde{m}}^{*}-1$
and $m=\tilde{m}$. Based on the threshold based policy, the source
should transmit one backlogged packet in each slot when the queue
length reaches the threshold $i_{m}^{*}$ (associated with the channel
state $m$) so as to minimize the average delay. As plotted in Fig.$\,$\ref{fig:threshold_strucutre},
the thresholds $\{i_{m}^{*}\}$ take the stair-step shape, which is
the optimum way to exploit the limited power resource over wireless
channels.

\subsection{{\normalsize{}Derivation of the Optimal Transmission Parameters}}

Based on the property of the optimal solution $\{y_{i,m}^{*}\}$ presented
in Theorem \ref{thm:optimal_solution_structure}, we will show how
to derive the optimal solution $\{y_{i,m}^{*}\}$ and determine the
optimal transmission parameters $\{g_{i,m}^{*}\}$ and $\{f_{i,m}^{*}\}$
thereafter. For ease of expression, we define two functions as: 
\begin{equation}
\varGamma(i)=\max_{i_{m}^{*}\leq i}\quad m=\begin{cases}
1, & 0=i_{1}^{*}\leq i<i_{2}^{*},\\
m, & i_{m}^{*}\leq i<i_{m+1}^{*},\\
M, & i\geq_{M}^{*},
\end{cases}
\end{equation}
\begin{equation}
\chi(k)=\frac{1-\sum_{m=1}^{k}\eta_{m}}{\xi\sum_{m=1}^{k}\eta_{m}}.\label{eq:chi_fun}
\end{equation}
We also define two series of probabilities $\{\varphi_{1,i}\}$ and
$\{\varphi_{2,i}\}$ ($i\in\{0,1,\cdots,Q\}$) as
\begin{equation}
\begin{split} & \varphi_{1,i}\\
= & \begin{cases}
0, & i=0,\\
\left(\chi(\varGamma(i-1))\right)^{i-i_{\varGamma(i-1)}^{*}}\prod\limits _{m=1}^{\varGamma(i-1)-1}\left(\chi(m)\right)^{i_{m+1}^{*}-i_{m}^{*}}, & i>0,
\end{cases}
\end{split}
\label{eq:varphi_p1}
\end{equation}
\begin{equation}
\begin{split} & \varphi_{2,i}\\
= & \begin{cases}
0, & i<i_{\tilde{m}}^{*},\\
-\frac{\eta_{\tilde{m}}}{\xi\sum\limits _{m=1}^{\tilde{m}-1}\eta_{m}}\left(\chi(\varGamma(i-1))\right)^{i-i_{\varGamma(i-1)}^{*}}\\
\quad\times\prod\limits _{m=\tilde{m}}^{\varGamma(i-1)-1}\left(\chi(m)\right)^{i_{m+1}^{*}-i_{m}^{*}}, & i\geq i_{\tilde{m}}^{*},
\end{cases}
\end{split}
\label{eq:varphi_p2}
\end{equation}
Based on the probabilities $\{\varphi_{1,i}\}$ and $\{\varphi_{2,i}\}$,
we can define two more parameters as 
\begin{equation}
\theta_{1}=\sum_{m=1}^{M}\eta_{m}\left[\varphi_{1,i_{m}^{*}}+(1+\xi)\sum_{i=i_{m}^{*}+1}^{i_{M}^{*}}\varphi_{1,i}\right]P_{m},\label{eq:theta_P1}
\end{equation}
\begin{equation}
\theta_{2}=\sum_{m=1}^{M}\eta_{m}(\varphi_{2,i_{m}^{*}}+(1+\xi)\sum_{i=i_{m}^{*}+1}^{i_{M}^{*}}\varphi_{2,i})P_{m}+\eta_{\tilde{m}}P_{\tilde{m}},\label{eq:theta_P2}
\end{equation}
which are linear combinations of the power levels $\{P_{m}\}$. 
\begin{thm}
\label{thm:optimal_pi_result} Let $\nu_{1}=\sum\limits _{i=0}^{i_{M}^{*}}\varphi_{1,i}$
and $\nu_{2}=\sum\limits _{i=0}^{i_{M}^{*}}\varphi_{2,i}$. The optimal
steady-state probability $\pi_{i}^{*}$ can be expressed as a linear
function of the optimal steady-state probability $\pi_{0}^{*}$: 
\begin{equation}
\pi_{i}^{*}=\frac{\varphi_{2,i}}{\nu_{2}}+\left(\varphi_{1,i}-\frac{\nu_{1}}{\nu_{2}}\varphi_{2,i}\right)\pi_{0}^{*},\label{eq:optimal_pi}
\end{equation}
where $\pi_{0}^{*}$ is obtained as 
\begin{equation}
\pi_{0}^{*}=\left(\frac{p_{max}}{\alpha}-\frac{\theta_{2}}{\nu_{2}}\right)\left(\theta_{1}-\frac{\nu_{1}}{\nu_{2}}\theta_{2}\right)^{-1}.\label{eq:optimal_pi0}
\end{equation}

\end{thm}
Once obtaining the steady-state probabilities $\{\pi_{i}^{*}\}$
given by (\ref{eq:optimal_pi}), we can compute the optimal solution
$\{y_{i,m}^{*}\}$ as 
\begin{equation}
y_{i,m}^{*}=\begin{cases}
0, & i<i_{m}^{*}-1,\\
\left(\sum\limits _{n=m}^{M}\frac{\eta_{n}}{\eta_{m}}\right)\pi_{i_{m}^{*}-1}^{*}-\left(\sum\limits _{n=1}^{m-1}\frac{\eta_{n}}{\eta_{m}}\right)\xi\pi_{i_{m}^{*}}^{*}, & i=i_{m}^{*}-1,\\
\pi_{i}^{*}+\xi\pi_{i+1}^{*}, & i\geq i_{m}^{*}.
\end{cases}\label{eq:optimal_yim}
\end{equation}
From its definition $y_{i,m}^{*}=\pi_{i}^{*}g_{i,m}^{*}+\xi\pi_{i+1}^{*}f_{i+1,m}^{*}$
(c.f. (\ref{eq:def_yim})), we can further determine the optimal transmission
probabilities $\{g_{i,m}^{*}\}$ and $\{f_{i,m}^{*}\}$. 
\begin{thm}
\label{thm:transmission_parameters}The pair of the optimal transmission
probabilities $g_{i,m}^{*}$ and $f_{i+1,m}^{*}$ satisfy
\begin{equation}
\begin{cases}
g_{i,m}^{*}=f_{i+1,m}^{*}=0, & i<i_{m}^{*}-1\\
g_{i,m}^{*}=f_{i+1,m}^{*}=1, & i\geq i_{m}^{*}
\end{cases}\label{eq:transmission_parameter_v1}
\end{equation}
for any channel state $m$ and for $i=i_{m}^{*}-1$ 
\begin{equation}
\begin{cases}
g_{i_{m}^{*}-1,m}^{*}=f_{i_{m}^{*},m}^{*}=0, & m\neq\tilde{m},\\
\pi_{i_{\tilde{m}}^{*}-1}^{*}g_{i_{\tilde{m}}^{*}-1,\tilde{m}}^{*}+\xi\pi_{i_{\tilde{m}}^{*}}^{*}f_{i_{\tilde{m}}^{*},\tilde{m}}^{*}=y_{i_{\tilde{m}}^{*}-1,\tilde{m}}^{*}, & m=\tilde{m}.
\end{cases}\label{eq:transmission_parameter_v2}
\end{equation}

\end{thm}
\emph{Remark}: In Theorem \ref{thm:transmission_parameters}, we
show that the delay optimal scheduling algorithm is a threshold-based
transmission scheme. When the channel state is $m$, the source transmits
one backlogged packet with power $P_{m}$ if the queue length reaches
the threshold $i_{m}^{*}$, and otherwise remains silent. Note that
the optimal threshold is $i_{1}^{*}=0$ when the channel state is
$m=1$. This implies that  the source transmits one data packet,
either newly arriving or stored, in each slot provided the channel
condition is the best. There may exist multiple pairs of transmission
probabilities $g_{i_{\tilde{m}}^{*}-1,\tilde{m}}^{*}$ and $f_{i_{\tilde{m}}^{*},\tilde{m}}^{*}$
which satisfy the equation (\ref{eq:transmission_parameter_v2}). 

It is not trivial to derive the integer thresholds $\{i_{m}^{*}\}$.
Fortunately, we are able to reduce the computational complexity significantly
by exploiting the monotonic property \emph{$0=i_{1}^{*}\leq\cdots\leq i_{M}^{*}$}.
For example, nested bisection methods can be applied to find the optimal
thresholds $\{i_{m}^{*}\}$. Then, we show that an elegant expression
of the optimal thresholds can be derived in the case with $M=2$. 
\begin{cor}
In the two-state channel case with $M=2$, the optimal transmission
probabilities can be expressed as $g_{i,1}^{*}=f_{i+1,1}^{*}=1$ for
all $i\geq0$, and 
\begin{equation}
\begin{cases}
g_{i,2}^{*}=f_{i+1,2}^{*}=0, & i<i_{2}^{*}-1,\\
\eta_{2}(\pi_{i}^{*}g_{i,2}^{*}+\xi\pi_{i+1}^{*}f_{i+1,2}^{*})=x_{i,2}^{*}, & i=i_{2}^{*}-1,\\
g_{i,2}^{*}=f_{i+1,2}^{*}=1, & i\geq i_{2}^{*},
\end{cases}
\end{equation}
where the threshold $i_{2}^{*}$ is given by
\begin{equation}
i_{2}^{*}=\begin{cases}
\lfloor\log_{\frac{1-\eta_{1}}{\eta_{1}\xi}}(1-\frac{1}{\pi_{0}^{*}}(1-\frac{1-\eta_{1}}{\eta_{1}\xi}))\rfloor, & \frac{1-\eta_{1}}{\eta_{1}\xi}\neq1,\\
\lfloor(\pi_{0}^{*})^{-1}\rfloor, & \frac{1-\eta_{1}}{\eta_{1}\xi}=1.
\end{cases}\label{eq:opt_threshold_i2}
\end{equation}

\end{cor}
One can see that this result is equivalent to that in \cite{2007_WChen_IWCLD}.
Therefore, the delay optimal scheduling policy proposed in \cite{2007_WChen_IWCLD}
is indeed the optimal for a two-state wireless channel.

\section{Simulation Results\label{sec:Simulation_Results}}

\noindent In this section, simulation results are presented to demonstrate
the performance of the proposed scheduling scheme and validate our
theoretical analysis. Each simulation runs over $10^{6}$ time slots.
In each slot, the packet transmissions are scheduled according to
our proposed policy. In the figures, the solid lines and the marks
'o' indicate theoretical and simulation results, respectively. One
can see that theoretical and simulation results match well. 
\begin{figure}[t]
\centering\includegraphics[width=0.45\textwidth]{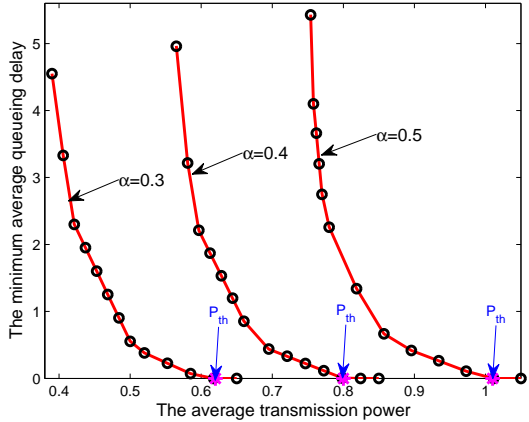}

\protect\caption{The delay-power tradeoff curves  for different data arrival rates
$\alpha$. \label{fig:delay_power_alpha}}
\end{figure}

Fig.$\,$\ref{fig:delay_power_alpha} plots the optimal delay-power
tradeoff performance for the data arrival rate $\alpha=0.3$, $0.4$
and $0.5$, respectively. We set $M=3$, $[\eta_{1}\,\eta_{2}\,\eta_{3}]=[0.25,\,0.5,\,0.25]$,
the buffer capacity $Q$ is set to $100$, and the transmission power
levels $\{P_{m}\}=\{1,2,3\}$ (watt). It is observed from Fig.$\,$\ref{fig:delay_power_alpha}
that the minimum average queueing delay monotonically decreases with
the increase of the power constraint $p_{max}$. When more power can
be consumed, the packets can be delivered more quickly since the source
does not necessarily wait for the channel quality to turn better.
 When the power resource is sufficient with $p_{max}\geq P_{th}$,
each newly arriving packet can be delivered immediately regardless
of the channel state and thus the queueing delay is zero. We also
notice that to achieve a same average delay $\bar{D}$, the source
consumes more power when data packets arrive more frequently with
larger $\alpha$. 

\begin{figure}[t]
\centering\includegraphics[width=0.45\textwidth]{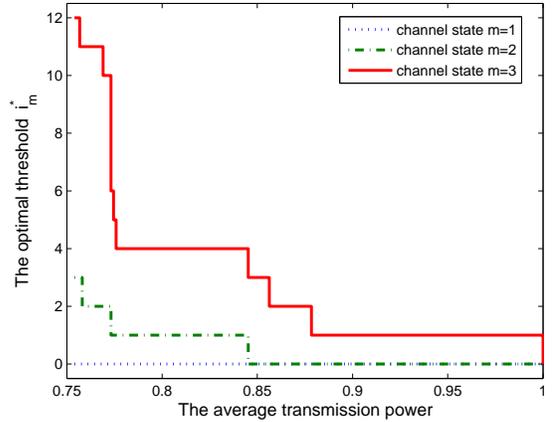}

\protect\caption{The optimal threshold $i_{m}^{*}$ for each channel state $m$. \label{fig:optimal_threshold}}
\end{figure}

In Fig.$\,$\ref{fig:optimal_threshold}, we plot the optimal threshold
on the queue length $i_{m}^{*}$ for the channel state $m=1$, $2$
and $3$, respectively, when the data arrival rate is $\alpha=0.5$.
The thresholds $\{i_{m}^{*}\}$ satisfy $i_{1}^{*}\leq i_{2}^{*}\leq i_{3}^{*}$
for any power constraint. This means that  the source should exploit
relatively better channel conditions to transmit as possible. When
the channel quality is the best with $m=1$, the optimal threshold
is equal to $i_{m}^{*}=0$, regardless of the average power constraint
$p_{max}$. When $m=2$ or $m=3$, the optimal threshold $i_{m}^{*}$
steps down with the increase of the power $p_{max}$. Meanwhile, a
smaller minimum average delay is observed from Fig.$\,$\ref{fig:delay_power_alpha},
since $\bar{D}^{*}=\alpha^{-1}\sum_{i=0}^{i_{M}^{*}}i\pi_{i}^{*}$
highly depends on the largest queue length $i_{M}^{*}$. When the
average power is sufficiently large, $i.e.$, $p_{max}\geq P_{th}$,
all the thresholds $\{i_{m}^{*}\}$ are equal to zero and the minimum
average delay is zero. 

\section{Conclusions\label{sec:Conclusions}}

In this paper, we investigated the delay optimal scheduling problem
over a $M$-state wireless channel with fixed modulation. For a target
BER, the source adjusts its transmission power $P_{m}$ according
to the channel state $m$. In this system, we proposed a stochastic
scheduling policy: the source transmits one packet with probability
$g_{i,m}$ or $f_{i,m}$, depending on whether there is new data arrival,
when the queue length is $i$ and the channel state is $m$. By Markov
chain modeling and variable substitution, we constructed an LP problem
to minimize the average delay under the average power constraint.
By exploiting the property of the LP problem, we revealed the structure
of the optimal solution and then derived the optimal probabilities
$\{g_{i,m}^{*}\}$ or $\{f_{i,m}^{*}\}$. It was found that the source
should always transmit as long as the channel quality is best. Otherwise,
the source holds from transmission when the queue length is below
the optimal threshold $i_{m}^{*}$, and transmits with power $P_{m}$
when the data queue length exceeds the threshold $i_{m}^{*}$, given
the channel state $m$. Simulation results confirmed our theoretical
analysis. 

\bibliographystyle{IEEEtran}

\end{document}